\documentclass[final,5p,times]{elsarticle}

\usepackage{amssymb}
\usepackage{lipsum}
\usepackage[french,english]{babel}
\usepackage[T1]{fontenc}
\usepackage{graphicx}
\usepackage{dcolumn}
\usepackage{bm}
\usepackage{hyperref}
\usepackage[version=4]{mhchem}
\usepackage{subcaption}
\usepackage{color}
\usepackage{tabularx}
\usepackage{siunitx}
\DeclareSIUnit\atm{atm}
\usepackage{enumerate}
\usepackage{array}
\journal{Solid State Ionics}

\begin{document}

\begin{frontmatter}



\title{Defect thermodynamics of orthorhombic Ba$_2$In$_2$O$_5$:\\ First-principles calculations on the role of oxygen dumbbell interstitials}


\author[first]{Rachele Sciotto}
\author[first]{Karsten Albe}
\affiliation[first]{organization={Department of Materials and Earth Sciences, Materials Modelling Division, Technical University of Darmstadt},
            addressline={Otto-Berndt-Str. 3}, 
            city={Darmstadt},
            postcode={64287}, 
            country={Germany}}

\begin{abstract}
The brownmillerite-type barium indate (\ce{Ba2In2O5}) is a potential electrolyte for mixed ionic-electronic conduction in solid oxide fuel cells. Revealing the defect chemistry of this material is key to understanding its ionic and electronic conductivity. 
In this contribution, we report the existence of oxygen interstitials in a dumbbell configuration, which are also observed in \ce{In2O3}.
Using Density Functional Theory within the generalized gradient approximation, complemented by selected hybrid-functional calculations, we investigate vacancies, various oxygen interstitials, and Frenkel pairs. In doing so, we evaluate the formation energies, charge transition levels, and concentrations as a function of oxygen partial pressure. Our results show that oxygen vacancies and interstitials dominate the intrinsic defect landscape. Among the interstitials, we identify stable dumbbell configurations that remain neutral across the entire band gap. Other interstitial configurations exhibit charged states and become the prevailing compensating defects at high oxygen partial pressures, alongside oxygen vacancies. Our results provide a consistent picture of the thermodynamics of intrinsic defects in barium indate, setting the stage for future investigations of the diffusion dynamics of oxygen vacancies and interstitials.
\end{abstract}




\end{frontmatter}




\section{Introduction}
\label{introduction}

Mixed ionic electronic conductors (MIECs) 
have been extensively studied for applications in energy storage, sensors, gas permeation membranes, and, most notably, solid oxide fuel cells (SOFCs) due to their ability to conduct ionic and electronic charge carriers simultaneously \cite{schmidtSafeDesignOperation2001,shaoPerformanceMixedconductingCeramic2001, wangPerovskiteHollowFiberMembranes2005, zhangSynthesisCharacterizationEvaluation2008,riessMixedIonicElectronic2003}.
A fuel cell directly converts the chemical energy of a fuel gas into electrical energy through electrochemical reactions between a fuel and an oxidant, without an intermediate combustion step. This enables higher conversion efficiencies than conventional combustion-based power generation. Unlike batteries, fuel cells do not require recharging, as they continuously generate electricity as long as fuel and oxidant are supplied \cite{stambouliSolidOxideFuel2002}.
    
Goodenough \textit{et al.} \cite{goodenoughOxideionConductionBa2In2O51990} identified barium indate (\ce{Ba2In2O5}, BIO) as a promising fast oxide-ion conductor due to its unique brownmillerite-type structure. This oxygen-deficient perovskite-related compound features a high concentration of oxygen vacancies, with one of every six oxygen sites missing. However, these vacancies are not randomly distributed; instead, they adopt a specific ordering that alternates between layers of \ce{InO6} octahedra and \ce{InO4} tetrahedra, which restricts their mobility.
At temperatures above \qty{925}{\celsius}, BIO undergoes an order-disorder transition, where previously ordered oxygen vacancies begin to randomize, leading to a sharp increase in ionic conductivity. This transition is accompanied by a structural phase change, first from orthorhombic ($Ibm2$, space group No. 46-2) to a tetragonal phase ($I4cm$, space group No. 108) between \qty{925}{\celsius} and \qty{1040}{\celsius}, and then to a cubic perovskite phase ($Pm\overline{3}m$, space group No. 221) above \qty{1040}{\celsius}, in which the oxygen vacancies are fully disordered \cite{speakmanInsituDiffractionStudy2002, leeInitioStudyOxygenVacancy2006}. Computational studies by Mohn \textit{et al.} \cite{mohnCollectiveIonicMotion2004, mohnOrderDisorderedState2005} have shown that the energetically preferred arrangement of oxygen vacancies in BIO is a staggered configuration rather than a stacked alignment.
    
The limited mobility of oxygen vacancies in BIO is reminiscent of that observed in bixbyite \ce{In2O3} (IO) \cite{agostonInitioModelingDiffusion2010}, suggesting that similar defect mechanisms might be at play in both materials.
A recent photoelectron spectroscopy study on the defect chemistry of IO and \ce{Sn}-doped \ce{In2O3} (ITO) by Klein \textit{et al.} \cite{kleinOriginQuantificationUltimate2024a} found that cathodic polarization of the material leads to oxygen removal and an increase in Fermi level, reaching a maximum value of $E_{\textrm{F}} - E_{\textrm{VB}} = \qty[input-uncertainty-signs]{3.85(10)}{\eV}$ for ITO and $E_{\textrm{F}} - E_{\textrm{VB}} = \qty[input-uncertainty-signs]{3.35(10)}{\eV}$ for IO. Further increasing the applied voltage leads to the reduction of \ce{Sn^{4+}} to \ce{Sn^{2+}}, suggesting that the ultimate limitation of carrier concentration in ITO is determined by reduction of the dopant rather than the self-compensating formation of interstitial oxygen atoms.
    
While the defect chemistry of IO is relatively well understood, much less is known about BIO. An upper limit to the Fermi level could potentially be set by the formation of oxygen interstitials, a possibility that remains largely unexplored. In moderately reducing environments, oxygen interstitials might also act as compensating defects, similar to their role in IO \cite{agostonInitioModelingDiffusion2010, kleinOriginQuantificationUltimate2024a}.
A defect model for BIO was first proposed by Zhang and Smyth \cite{zhangDefectsTransportBrownmillerite1995}. Their results show that the main defects are intrinsic anion Frenkel defects below the order-disorder transition temperature of \qty{925}{\celsius}, and that above it, the oxide can be treated as acceptor-doped perovskite with extrinsic oxygen vacancies. 
Another subsequent study by Fisher and Islam \cite{fisherDefectProtonsConductivity1999} applied simulation techniques based on the Born model of polar solids to investigate the defect energetics of BIO. They also concluded that oxygen Frenkel pairs are the dominant intrinsic defects in the low-temperature phase, with a preference for oxide ion diffusion via these defects in the [001] direction.
However, both of these defect models consider only oxygen interstitials filling the structural oxygen vacancy sites in the tetrahedral layer.
    
In this work, we use density functional theory to clarify the role of intrinsic defects, with a focus on oxygen interstitials, in orthorhombic BIO from first principles. We investigate defect formation energies and charge transition levels of vacancies, oxygen interstitials, and Frenkel pairs in BIO. 
By analyzing the defect thermodynamics across different chemical potential conditions and experimental oxygen partial pressures, we establish the accessible range of Fermi levels and the charge compensation mechanisms in BIO. 
This provides new insights into the defect chemistry of BIO, offering guidance for doping engineering of brownmillerite-type materials.
    
This paper is structured as follows: Section \ref{sec:methods} outlines the computational methodology, Section \ref{sec:res_disc} discusses intrinsic defects, and Section \ref{sec:conc} presents the conclusions and prospects for future work.

\section{Methodology}
\label{sec:methods}
\subsection{Computational details}
\label{sec:comp_det}
Density functional theory (DFT) calculations were performed using the Vienna Ab Initio Simulation Package (VASP) \cite{kresseInitioMolecularDynamics1993,kresseEfficiencyAbinitioTotal1996,kresseEfficientIterativeSchemes1996}. The projector-augmented wave (PAW) scheme \cite{blochlProjectorAugmentedwaveMethod1994, kresseUltrasoftPseudopotentialsProjector1999} described the electron-ion interaction.  For the valence electron configuration, \ce{Ba} was considered with $5s^2~5p^6~5d^{0.01}~6s^{1.99}$, \ce{In} with $4d^{10}~5s^2~5p^1$, and \ce{O} with $2s^2~2p^4$.

\begin{figure}[t]
    \centering
    \includegraphics[width=1\linewidth]{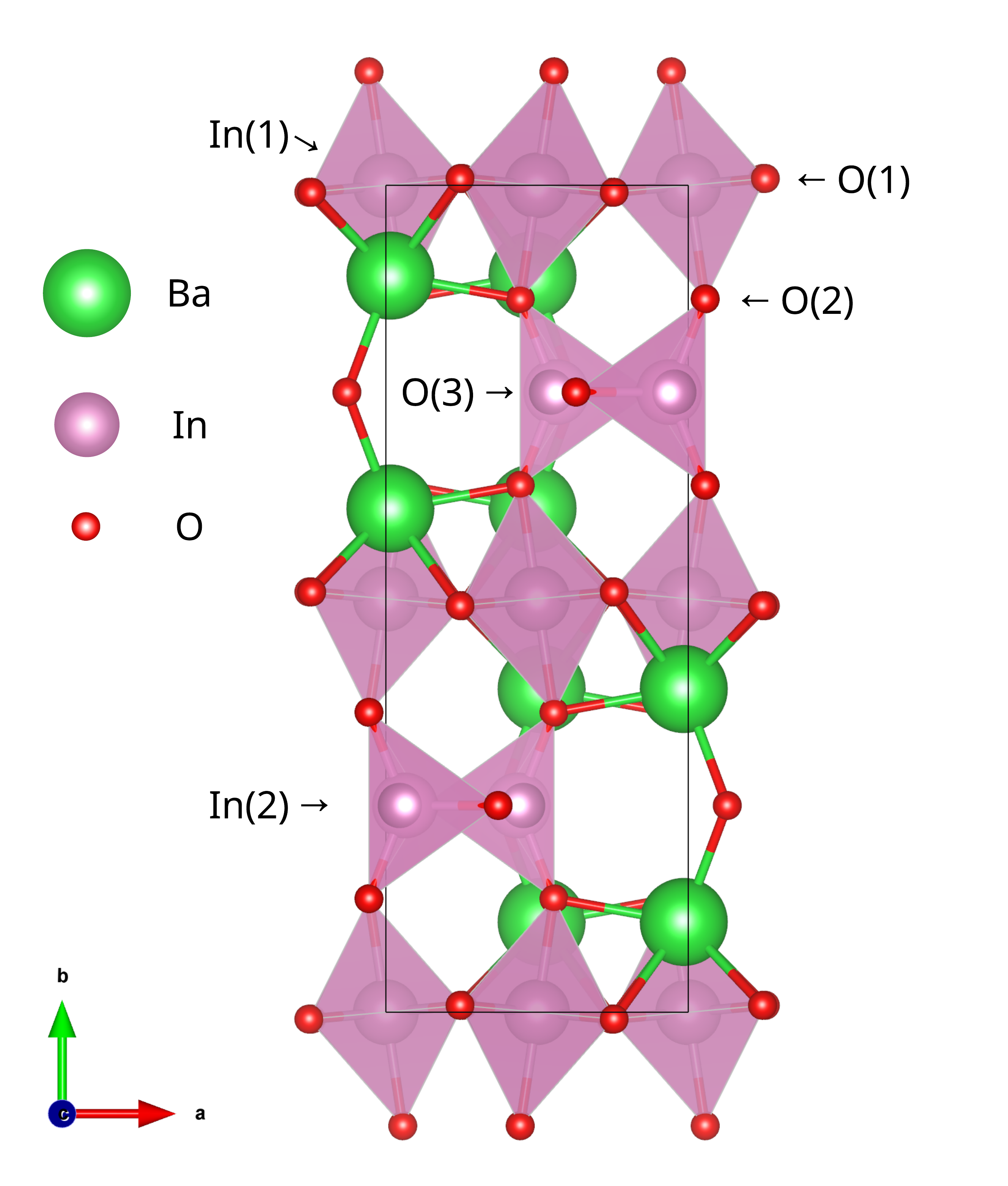}
    \caption{Crystal structure of orthorhombic BIO with 36 atoms. Ba, In, and O atoms are identified in green, pink, and red, respectively. The three distinct oxygen sites and the two distinct indium sites are marked.}
    \label{fig:BIO_crystal}
\end{figure}

Hybrid functional calculations were used exclusively to obtain an accurate reference electronic structure for BIO. Specifically, the density of states (DOS) and the band structure of the room-temperature orthorhombic phase ($Ibm2$, space group No. 46-2) were calculated using the Heyd-Scuseria-Ernzerhof (HSE06) functional \cite{heydHybridFunctionalsBased2003, heydErratumHybridFunctionals2006, krukauInfluenceExchangeScreening2006}, with a mixing of 35\% and a screening parameter of 0.2 \AA$^{-1}$, employing the 36-atom cell shown in Figure \ref{fig:BIO_crystal}. The wave functions were expanded into plane waves up to an energy cutoff of 650 eV. 
The Brillouin zone was sampled using a $\Gamma$-centered $4 \times 2 \times 4$ $k$ mesh. The atomic positions were relaxed until the Hellmann-Feynman forces acting on each atom were below 0.05 eV/Å.

Given the high number of calculations required to study intrinsic defects, including defect complexes, using hybrid functionals would be computationally expensive. Therefore, for these calculations, the electron exchange and correlation effects for the defect calculations were treated within the generalized gradient approximation (GGA) using the Perdew-Burke-Ernzerhof functional (PBE) \cite{perdew_generalized_1996} for structural relaxation. To correct for the well-known underestimation of the band gap by GGA functionals, the band gap was corrected using the method proposed in Refs. \cite{perssonTypeDopingCuIn2005, lanyAssessmentCorrectionMethods2008}, with HSE06 providing the reference band edges and the band gap.

All defect calculations were performed using a $2\times 1 \times 2$ supercell containing 144 atoms. The same plane-wave cutoff and relaxation criterion were used as for the HSE06 reference calculations; however, the Brillouin zone was sampled using a $\Gamma$-centered $2 \times 2 \times 2$ $k$ mesh for the supercell.

The Voronoi interstitial generator provided by the Python Materials Genomics (\textsc{pymatgen}) library \cite{ongLiFePO2PhaseDiagram2008, ongThermalStabilitiesDelithiated2010} was used to identify potential interstitial sites.
Structural models were visualized using VESTA \cite{mommaVESTA3Threedimensional2011}, and figures were prepared with the Python packages \texttt{pynter} and \texttt{defermi} \cite{villaImpactDopingConditions2023a, defermi}.

\subsection{Defect thermodynamics}
\label{sec:def_thermo}
The formation energy $\Delta E_{D}^f$ of a defect $D$ with charge $q$ can be written as \cite{freysoldt_first-principles_2014}

\begin{multline}
    \label{eq:form_en}
    \Delta E_{D}^f = E_{D} - E_{P} + q(E_{\textrm{VBM}} + \mu_e) + E_{\textrm{corr}} - \sum_i \Delta n_i \mu_i,
\end{multline}

where $E_{D}$ is the total energy of the system with defect $D$ and charge $q$, and $E_{P}$ is the energy of the pristine reference cell. The third term describes the dependence of the formation energy on the chemical potential of the electrons $\mu_e$, while the position of the maximum valence band (VBM) $E_{\textrm{VBM}}$ is taken as a reference. $E_{\textrm{corr}}$ describes the applied energy corrections. The spurious electrostatic interactions between periodic images of charged defects, which arise due to the finite size of the supercell, are corrected using the scheme proposed by Kumagai \textit{et al.} \cite{kumagai_electrostatics-based_2014}. It is well known that semi-local functionals, such as PBE, severely underestimate the band gap and can, in some cases, introduce uncertainties in calculated defect energetics \cite{freysoldt_first-principles_2014}. To account for this, the band gap is corrected by rigidly shifting the valence band maximum (VBM) and the conduction band minimum (CBM) as obtained from PBE calculations to the respective values obtained by HSE06 calculations \cite{perssonTypeDopingCuIn2005, lanyAssessmentCorrectionMethods2008}. The correction energy is then given by
    
\begin{equation}
    \label{eq:bg_corr}
    \Delta E_{bg} = n_e\Delta E_{\mathrm{CBM}} + n_h\Delta E_{\mathrm{VBM}},
\end{equation}

where $n_e$ and $n_h$ are the number of electrons introduced in the conduction band and the number of holes introduced in the valence band, respectively. $\Delta E_{\mathrm{CBM}}$ and $\Delta E_{\mathrm{VBM}}$ are the shifts in the CBM and VBM between HSE06 and PBE. Here, donor defects follow the energy correction of the CBM, while acceptor defects follow the correction of the VBM. Since the VBM is shifted downward, $E_\mathrm{VBM}$ in Equation \ref{eq:form_en} is also shifted by

\begin{equation}
    E_\mathrm{VBM} = E_\mathrm{VBM}^{\mathrm{PBE}}-\Delta E_{\mathrm{VBM}}.
\end{equation}

Finally, the last term accounts for the dependence of the formation energy on the chemical potentials of the constituent elements. Here, $\Delta n_i$ is the difference between the number of atoms of species $i$ in the defective and bulk systems. The chemical potential $\mu_i$ of element $i$ can be rewritten as

\begin{equation}
    \mu_i=\mu_i^{\textrm{ref}}+\Delta\mu_i,
\end{equation}

where $\mu_i^{\textrm{ref}}$ is the chemical potential of the element $i$ in its stable elemental phase.
The DFT data used to generate the phase stability diagram for the orthorhombic brownmillerite BIO were obtained from the Materials Project (MP) database (\textsc{pymatgen} implementation, version 2025.5.2) \cite{jainCommentaryMaterialsProject2013, ongLiFePO2PhaseDiagram2008,ongThermalStabilitiesDelithiated2010}, which uses consistent computational parameters with respect to our defect calculations. 
    
A charge transition level (CTL) is the electron chemical potential at which the formation energy of a given defect in two different charge states ($q_1$ and $q_2$) becomes equal:

\begin{equation}
    \label{eq:ctls}
    \varepsilon = - \frac{\Delta E_D^F(q_1, \mu_e=0)-\Delta E_D^F(q_2, \mu_e=0)}{q_1-q_2}.
\end{equation}

Here, $\Delta E_D^F(q_1, \mu_e=0)$ and $\Delta E_D^F(q_2, \mu_e=0)$ are the defect formation energies at charge states $q_1$ and $q_2$ respectively, when the electron chemical potential $\mu_e$ is at the VBM.
At thermodynamic equilibrium, the concentration of a defect $D$ in charge state $q$ is given by

\begin{equation}
    [D_q] = N_{\textrm{sites}}\exp{\left(- \frac{\Delta E_{q,D}^F(\mu_e)}{k_B T}\right)},
\end{equation}

where $q$ represents the defect charge, $N_{\textrm{sites}}$ the concentration of the corresponding lattice sites, and $k_B$ the Boltzmann constant.
It is important to underline that the electron chemical potential $\mu_e$ is not a free parameter. It is rather determined by the charge neutrality condition

\begin{equation}
    \label{eq:charge_neutr}
    n_e + n_A = n_h + n_D,
\end{equation}

in which $n_e$ and $n_h$ are the electron and hole carrier concentrations, respectively, while $n_D$ and $n_A$ are the donor and acceptor concentrations, respectively.
Considering all defect species $D$ in their different charge states $q$, Equation \ref{eq:charge_neutr} can be reformulated as

\begin{equation}
    \sum_{q,D}{q\cdot[D_q]}-n_e+n_h=0.
\end{equation}

Intrinsic carrier concentrations are computed by integrating the number of unoccupied states up to the VBM ($E_\textrm{VBM}$) for holes,

\begin{equation}
    \label{eq:n_h}
    n_h = \int_{-\infty}^{E_\textrm{VBM}}{D(\varepsilon)[1-f(\varepsilon)]\textrm{d}\varepsilon},
\end{equation}

and the number of occupied states from the conduction band minimum (CBM) ($E_\textrm{CBM}$) for electrons,

\begin{equation}
    \label{eq:n_e}
    n_e = \int_{E_\textrm{CBM}}^{+\infty}{D(\varepsilon)f(\varepsilon)\textrm{d}\varepsilon},
\end{equation}

where $f(\varepsilon, \mu_e)=[\exp(\varepsilon-\mu_e)/k_B T]^{-1}$ is the Fermi-Dirac distribution.
    
In this study, the integrals in Equation \ref{eq:n_h} and \ref{eq:n_e} were computed as Riemann sums over the energies in the calculated density of states, following the method delineated by \textsc{pymatgen} \cite{ongLiFePO2PhaseDiagram2008, ongThermalStabilitiesDelithiated2010}. Subsequently, the equilibrium electron chemical potential $\mu_e$ was determined self-consistently by solving the charge neutrality condition for $\mu_e$ at a given temperature.
    
Equation \ref{eq:form_en} can also be applied to defect complexes. For such complexes, an important quantity is the binding energy $E_b$, defined as the difference between the formation energy of the complex $C$ and the sum of the formation energy of its individual constituent defects $D$:
    
\begin{equation}
    \label{eq:binding-energy}
    E_b = \Delta E_C^f -\sum_D{\Delta E_D^f}.
\end{equation}

This quantity is independent of chemical potentials and reflects the energetic cost or benefit associated with defect pairing. 
A negative binding energy indicates that the formation of isolated defects is energetically less favorable than the formation of the complex. In this case, the interaction between defects is attractive, and complex formation is thermodynamically preferred.

\subsection{Oxygen chemical potential}
\label{sec:O_chem_pot}
Defect formation energies depend on the chemical potential of the reservoir. For oxygen, $\mu_{\ce{O}}$ varies with its partial pressure $p_{\ce{O2}}$ and can be expressed via the ideal gas law as

\begin{equation}
    \label{eq:mu_O}
    \mu_{\ce{O}}(T, p_{\ce{O2}}) = \mu_{\ce{O}}(T,p^0)+\frac{1}{2}k_BT\ln\left( \frac{p_{\ce{O2}}}{p^0}\right).
\end{equation}

Here, $k_B$ is the Boltzmann constant, $T$ the temperature, $\mu_{\ce{O}}$ the oxygen chemical potential, and $p^0$ the reference pressure.
The energy calculated for the $\ce{O2}$ molecule at 0 K $\mu_{\ce{O}}(\textrm{0 K}, p^0)$ serves as the reference. The temperature dependence $\mu_{\ce{O}}(T,p^0)$ was calculated following Reuter and Scheffler's methodology \cite{reuterCompositionStructureStability2001}, using the thermochemical tables in \cite{stullJANAFThermochemicalTables1971}. This procedure enables a direct comparison between \textit{ab initio} and experimental results, since the oxygen chemical potential can be calculated from experimental temperature and partial pressure. Consequently, the defect concentration and the electron chemical potential as functions of the partial pressure of oxygen are accessible within the same theoretical framework.

\section{Results and discussion}
\label{sec:res_disc}
\subsection{Electronic structure}
Preliminary steps for defect thermodynamics calculations include optimizing the bulk structure and evaluating its electronic structure. To determine the defect formation energy, the total energy of the bulk supercell and the VBM are required. On the other hand, the density of states (DOS) is used in the computation of electron and hole concentrations, as shown in Equations \ref{eq:n_h} and \ref{eq:n_e}. The computed values of lattice parameters and energy gap are reported and compared with literature in Table \ref{tab:lattice_bandgap}.

\begin{table}
    \renewcommand{\arraystretch}{1.3} 
    \centering
    \caption{Values of lattice constants $a$, $b$, $c$ (\AA) and energy gap $E_g$ (eV) of orthorhombic BIO (space group $Ibm$2, No. 46-2, unless otherwise indicated), computed with the PBE \cite{perdew_generalized_1996} and HSE06 \cite{heydHybridFunctionalsBased2003, heydErratumHybridFunctionals2006, krukauInfluenceExchangeScreening2006} exchange-correlation functionals and confronted with computational data by Rasim \cite{rasim2011} and Yoon \textit{et al.} \cite{yoonPhotocatalyticCO2Reduction2018b}, the Materials Project database \cite{ongLiFePO2PhaseDiagram2008, ongThermalStabilitiesDelithiated2010} and experimental data \cite{widenmeyerEngineeringOxygenPathways2020, yoonPhotocatalyticCO2Reduction2018b, vlasovBandGapEngineering2019}.}
        \begin{tabular}{l c c c c}
        \hline
        \hline
           & $a$ (\AA) & $b$ (\AA) & $c$ (\AA) & $E_g$ (eV) \\
          \hline
          PBE & 6.15 & 16.81 & 6.00 & 0.90\\
          Ref (PBE, $Ima$2) \cite{rasim2011} & 6.24 & 6.06 & 16.88 & 0.96 \cite{yoonPhotocatalyticCO2Reduction2018b} \\
          MP (PBE, $Ima$2) \cite{ongLiFePO2PhaseDiagram2008, ongThermalStabilitiesDelithiated2010} & 6.00 & 6.15 & 16.81 & 0.92 \\
          HSE06 & 6.12 & 16.67 & 5.97 & 2.82\\
          Exp. \cite{widenmeyerEngineeringOxygenPathways2020} & 6.10 & 16.74 & 5.96 & 2.94 \cite{vlasovBandGapEngineering2019}\\
          \hline\hline
        \end{tabular}
    \label{tab:lattice_bandgap}
\end{table}

\begin{figure}
    \centering
    \includegraphics[width=1\linewidth]{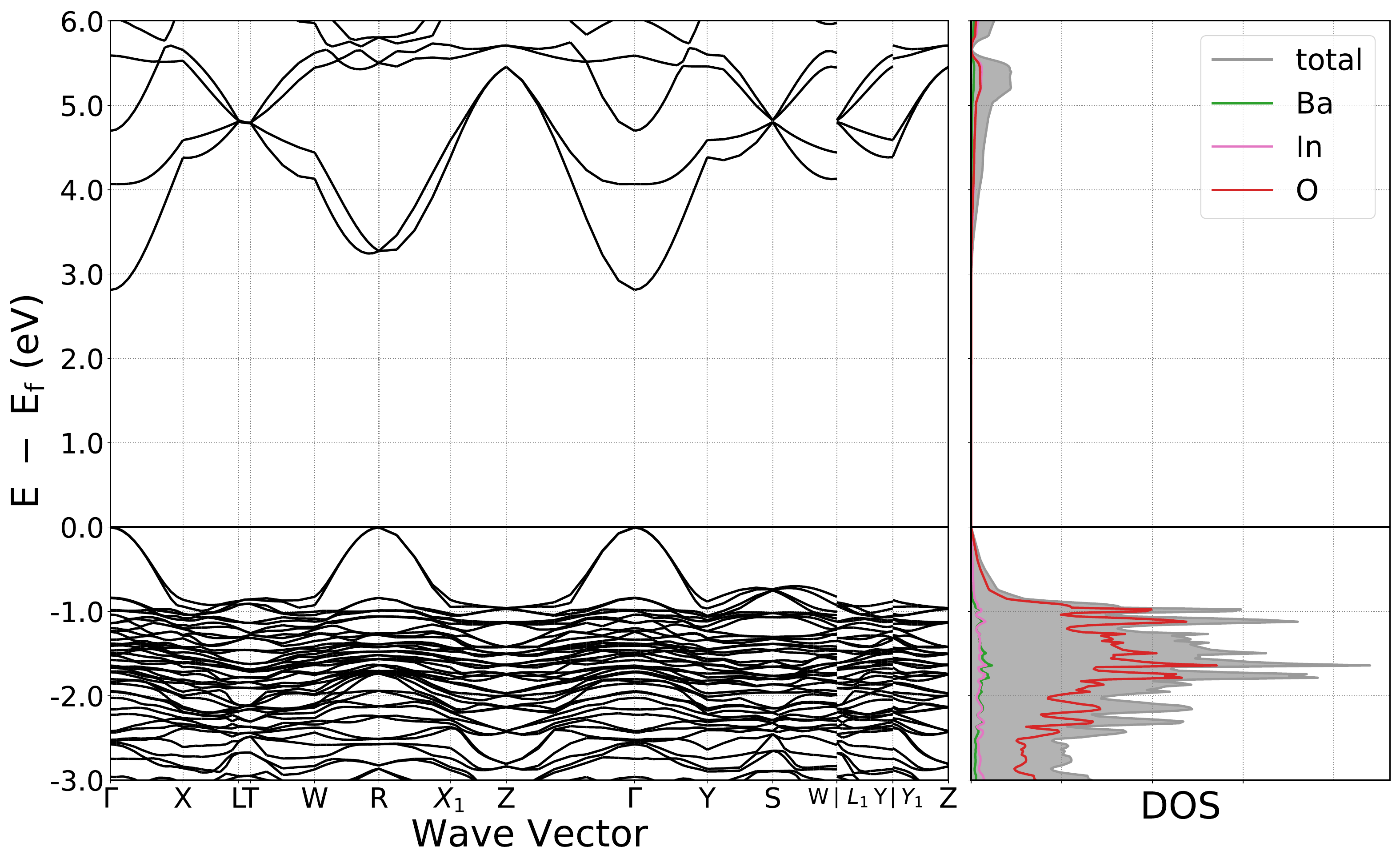}
    \caption{Band structure and density of states of orthorhombic BIO (space group \textit{Ibm}2) computed with the HSE06 functional.}
    \label{fig:BS-DOS}
\end{figure}

At the PBE level, we obtain a direct band gap at the $\Gamma$ point of 0.90 eV, significantly smaller than the experimental value of 2.94 eV \cite{vlasovBandGapEngineering2019}. It is well known that semilocal functionals, such as PBE, severely underestimate the band gap, and can, in some cases, introduce uncertainties in calculated defect energetics \cite{freysoldt_first-principles_2014}. To reduce this band-gap error, we therefore performed calculations using the HSE06 functional, which yields a gap of 2.82 eV, in closer agreement with experiment. The corresponding band structure and density of states (DOS) are shown in Figure \ref{fig:BS-DOS}. The valence band edge is mainly given by O $p$ states, while the conduction band is composed of O $p$ and In $d$ states.  

\subsection{Stability diagram}
\begin{figure*}
    \centering
    \includegraphics[width=0.8\textwidth]{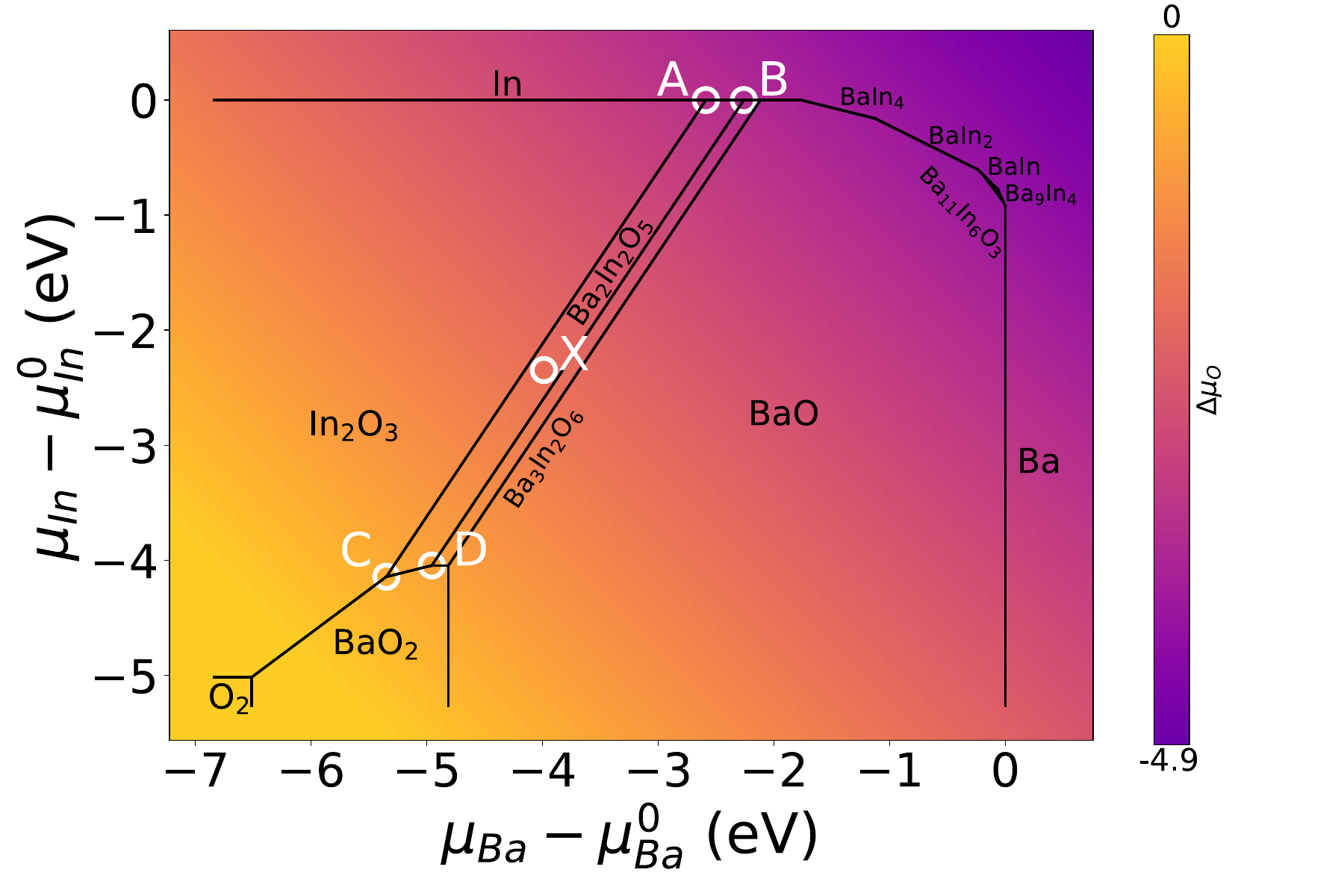}
    \caption{Stability diagram of the ternary Ba-In-O system, as derived from the data in Figure \ref{tab:pd_form_en}. The defect formation energies are discussed in terms of the chemical potentials at points A to D and point X as reported in Table \ref{tab:mu_values}.}
    \label{fig:sd}
\end{figure*}

To define the thermodynamic reservoirs, we construct the semigrand canonical phase diagram, which maps the stability regions as a function of the chemical potentials of the constituents. The list of stable phases is reported in Table \ref{tab:pd_form_en}. The resulting stability diagram, obtained from these formation energies, is shown in Figure \ref{fig:sd}. The allowed range of chemical potential is constrained by the condition that their sum must be equal to the formation enthalpy of BIO:

\begin{equation}
    \label{eq:delta_mu}
    2\Delta\mu_{\ce{Ba}} + 2\Delta\mu_{\ce{In}} + 5\Delta\mu_{\ce{O}} = \Delta H_f[\ce{Ba2In2O5}]
\end{equation}

For a fixed sum of $\Delta\mu_{\ce{Ba}} + \Delta\mu_{\ce{In}}$, the oxygen chemical potential $\Delta\mu_{\ce{O}}$ is determined by this condition. The stability region of BIO is further limited by the presence of competing phases and is represented by the area delimited by points A to D in Figure \ref{fig:sd}. More specifically, given a generic competing phase of composition \ce{Ba_iIn_jO_k}, the chemical potentials have to satisfy the condition:

\begin{equation}
    i\Delta\mu_{\ce{Ba}} + j\Delta\mu_{\ce{In}} + k\Delta\mu_{\ce{O}} \leq \Delta H_f[\ce{Ba_iIn_jO_k}].
\end{equation}

The defect formation energies were calculated for different thermodynamic reservoirs, represented in Figure \ref{fig:sd} by the labels from A to D and X. Regions A-D span from O-rich conditions (points C and D) to Ba-rich and In-rich conditions (points A and B).
We also evaluate the formation energies using a thermodynamic reservoir that represents the experimental conditions. In particular, we fix the oxygen chemical potential ($\Delta\mu_{\ce{O}}$) to its value under the conditions of temperature and partial pressure of the sintering process, according to the work of Chaoudhary \textit{et al.} \cite{chaoudharyUncoveringElectronicEffects2025} ($T=\qty{1573}{K}, p_{\ce{O2}}=\qty{0.21}{atm}$). Equation \ref{eq:mu_O}, described in Section \ref{sec:O_chem_pot}, details the calculation of this quantity. With this procedure, we obtain $\Delta\mu_{\ce{O}}=\qty{-1.85}{eV}$, which is indicated by the point X in Figure \ref{fig:sd}.
Since the range of stability varies for every compound, we have evaluated the chemical potential of the elements in O-poor and O-rich regions, assigning the former to reservoirs A and B and the latter to reservoirs C and D. In region X, the values are taken from the center of the stability region, having fixed $\Delta\mu_{\ce{O}}$ to $\qty{-1.85}{eV}$. This procedure is employed to compare the formation energies of intrinsic defects. The chemical potential values associated with the labeled regions in the stability diagram are reported in Table \ref{tab:mu_values}.
    
\begin{table}[]
    \renewcommand{\arraystretch}{1.3} 
    \centering
    \caption{Formation energy per atom for all the materials used to compute the stability diagram.}
        \begin{tabular*}{\columnwidth}
        {@{\extracolsep{\fill}}l c c @{}}
        \hline
        \hline
        Composition    & Structure                                        & $\Delta E^f$ (eV/atom)  \\
        \hline
        \ce{BaO2}      & Tetragonal (\textit{I}4\textit{/mmm})            & -2.17 \\
        \ce{Ba9In4}    & Tetragonal (\textit{I}4\textit{/m})              & -0.28 \\
        \ce{In}        & Trigonal (\textit{R}$\overline{3}$\textit{m})      & 0     \\
        \ce{Ba}        & Cubic (\textit{Im}$\overline{3}$\textit{m})        & 0     \\
        \ce{O2}        & Monoclinic (\textit{C}12\textit{/m}1)            & 0     \\
        \ce{BaO}       & Cubic (\textit{Fm}$\overline{3}$\textit{m})       & -2.82 \\
        \ce{Ba3In2O6}  & Tetragonal (\textit{I}4\textit{/mmm})            & -2.50 \\
        \ce{Ba2In2O5}  & Orthorhombic (\textit{Ibm}2)                     & -2.42 \\
        \ce{BaIn2}     & Orthorhombic (\textit{Imma})                     & -0.48 \\
        \ce{In2O3}     & Cubic (\textit{Ia}$\overline{3}$)                  & -1.99 \\
        \ce{BaIn4}     & Tetragonal (\textit{I}4\textit{/mmm})            & -0.35 \\
        \ce{BaIn}      & Orthorhombic (\textit{Cmcm})                     & -0.42 \\
        \ce{Ba11In6O3} & Tetragonal (\textit{I}4\textit{/mcm})            & -1.12 \\
        \hline
        \hline
        \end{tabular*}
    \label{tab:pd_form_en}
\end{table}

\begin{table}
    \centering
    \caption{Values of the chemical potentials of the individual elements for each of the labeled regions in the stability diagram (Figure \ref{fig:sd}).}
    \renewcommand{\arraystretch}{1.3}%
        \begin{tabular*}{\columnwidth}{@{\extracolsep{\fill}}l c c c@{}}
        \hline\hline
        Reservoir & $\Delta\mu_{\ce{Ba}}$ & $\Delta\mu_{\ce{In}}$ & $\Delta\mu_{\ce{O}}$  \\
        \hline
        A & -2.59 & 0.00 & -3.34 \\
        B & -2.26 & 0.00 & -3.47 \\
        C & -5.35 & -4.14 & -0.58 \\
        D & -4.96 & -4.04 & -0.78 \\
        X & -3.99 & -2.35 & -1.85 \\
        \hline\hline
        \end{tabular*}
    \label{tab:mu_values}
\end{table}

\subsection{Oxygen interstitials}
\label{sec:O_int}
\begin{figure*}
    \centering
    \includegraphics[width=0.85\textwidth]{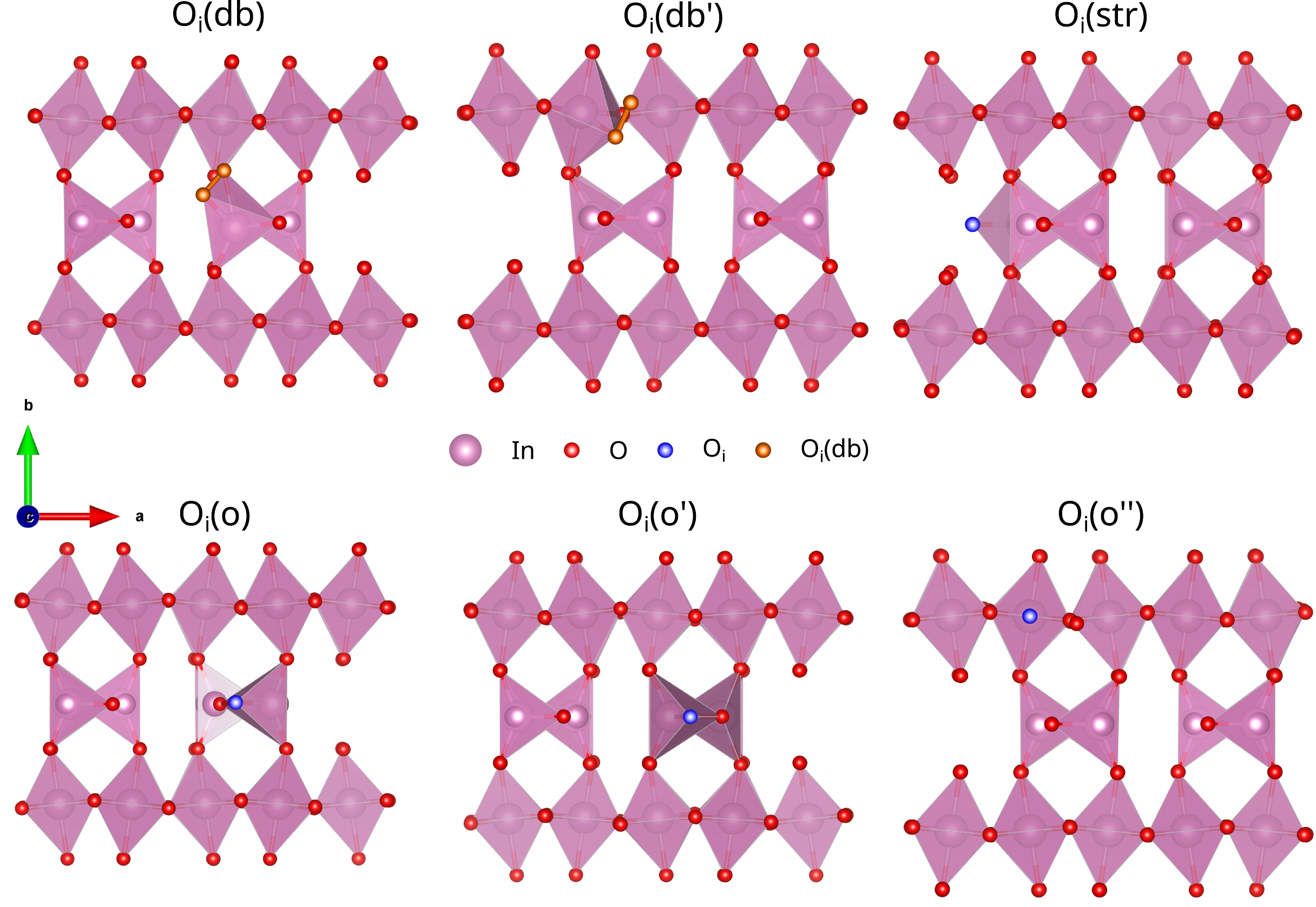}
    \caption{Crystal structure of orthorhombic BIO with the distinct oxygen interstitials. In, O, O$_i$, and O$_i$(db) atoms are identified in pink, red, blue, and orange, respectively.}
    \label{fig:distinct_Oi}
\end{figure*}

The Voronoi interstitial generator mentioned in Section \ref{sec:comp_det} identified eleven possible oxygen interstitial positions. Upon structural optimization, several of these starting configurations converged to the same local minima. In particular, six initial guesses relaxed into three identical sites; these duplicates were therefore removed, reducing the number of interstitial configurations to eight. 
Among the remaining eight, several configurations were still geometrically distinct but symmetrically equivalent within the BIO lattice, resulting in identical formation energies. Therefore, only one representative from each equivalent configuration was retained, leading to a final set of six distinct configurations.
The fractional coordinates of these symmetry-inequivalent interstitials are listed in Table \ref{tab:int_sites}, and their positions within the supercell are illustrated in Figure \ref{fig:distinct_Oi}. For completeness, their formation energies are shown in Figure S3 of the SM \cite{supplemental}. 

Among six unique configurations, two play a dominant role in the defect chemistry of BIO due to their lower formation energies:
\begin{enumerate}[(i)]
    \item O$_i$(db), a dumbbell-like O-O configuration located on the \ce{InO4} tetrahedron, and
    \item O$_i$(str), an interstitial occupying the structural oxygen vacancy site in the tetrahedral layer.
\end{enumerate}
    
The dumbbell configuration has also been reported in other materials, such as \ce{ZnO} and \ce{In2O3} \cite{erhartFirstprinciplesStudyStructure2005, erhartFirstprinciplesStudyIntrinsic2006, agostonInitioModelingDiffusion2010}. 

\subsection{Defect formation energies}

We have calculated defect formation energies according to Equation \ref{eq:form_en}, in the different regions of the stability diagram reported in Figure \ref{fig:sd} and Table \ref{tab:pd_form_en}. Due to the finite size of the supercell, we have corrected the electrostatic interactions between charged defect replicas using the method proposed by Kumagai \textit{et al.} \cite{kumagai_electrostatics-based_2014}.
We evaluated multiple charge states for each defect to determine the thermodynamically stable charge configurations and the associated charge transition levels (CTLs), as described by Equation \ref{eq:ctls}, to properly represent the contribution of each defect species to the charge neutrality condition. 
    
\begin{figure*}
    \centering
    \includegraphics[width=\textwidth]{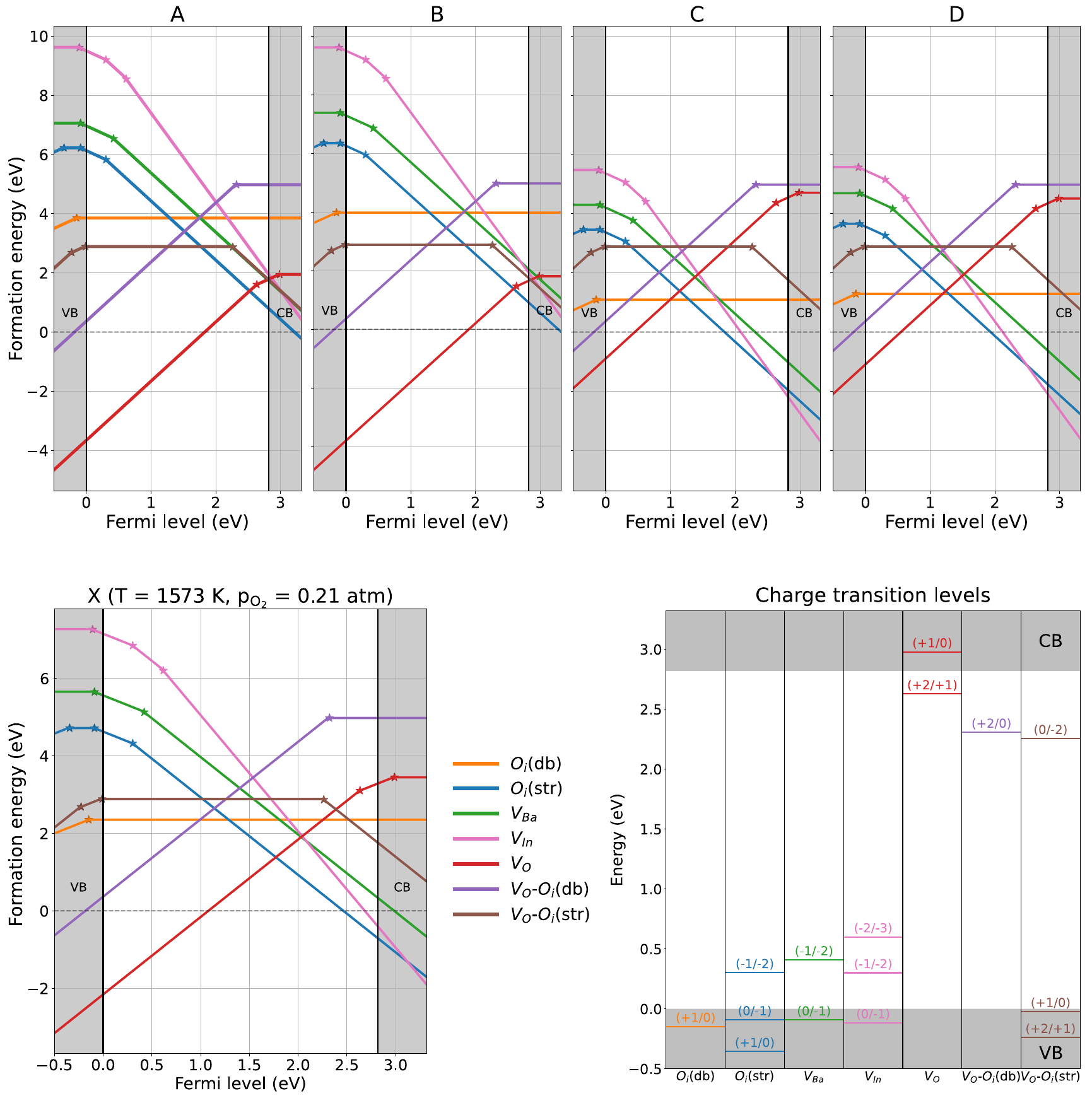}
    \caption{Defect formation energies as a function of the Fermi level position for the thermodynamic conditions indicated in Figure \ref{fig:sd} as A to D (top) and X (bottom left). The stars on the formation energy curves represent charge transition levels (bottom right).}
    \label{fig:Ef_ABCDX_CTLs}
\end{figure*}

In terms of intrinsic defects, we have studied the formation energies of the Ba, In and O vacancies (V$_{\ce{Ba}}$, V$_{\ce{In}}$, V$_{\ce{O}}$), the O interstitials (O$_{{i}}$), and the complex formed by an O interstitial and an O vacancy, also called a Frenkel pair (V$_{\ce{O}}$-O$_{{i}}$).
BIO contains three distinct crystallographic oxygen sites: O(1) in the equatorial plane of the \ce{InO6} octahedra, O(2) at the apical positions of the octahedra, and O(3) in the \ce{InO4} tetrahedra that are not shared with the octahedra. There are also two crystallographically distinct indium sites: In(1) in octahedral coordination, and In(2) in tetrahedral coordination. Barium atoms occupy a single crystallographic site.
Among the O and In vacancies, those at O(1) and In(2) exhibit the lowest formation energies (see Figure S1 and S2 in the SM \cite{supplemental}), and are therefore the focus of the discussion along with the single type of Ba vacancies. While the O(1) vacancy in the +2 charge state is the most stable configuration for Fermi levels up to 0.5 eV, the neutral O(2) vacancy becomes more stable at higher Fermi levels. Nevertheless, we focus on O(1) because of its role as a charge carrier.
For the sake of readability, we have removed the numerical index of the defects with multiple crystallographic sites: V$_{\ce{O}}$(1) and V$_{\ce{In}}$(2) will be referred to simply as V$_{\ce{O}}$ and V$_{\ce{In}}$, respectively.
As far as the oxygen interstitials are concerned, following our analysis delineated in Section \ref{sec:O_int}, we only consider O$_i$(db) and  O$_i$(str), i.e., the two interstitials with the lowest formation energy.
The formation energies computed in the boundary regions of the stability diagram are plotted as a function of the Fermi level in the top panel of Figure \ref{fig:Ef_ABCDX_CTLs}. The stars in the plot represent the charge transition levels, calculated according to Equation \ref{eq:ctls} and shown in the bottom right panel of Figure \ref{fig:Ef_ABCDX_CTLs}.

\begin{figure}[t]
    \centering
    \includegraphics[width=\linewidth]{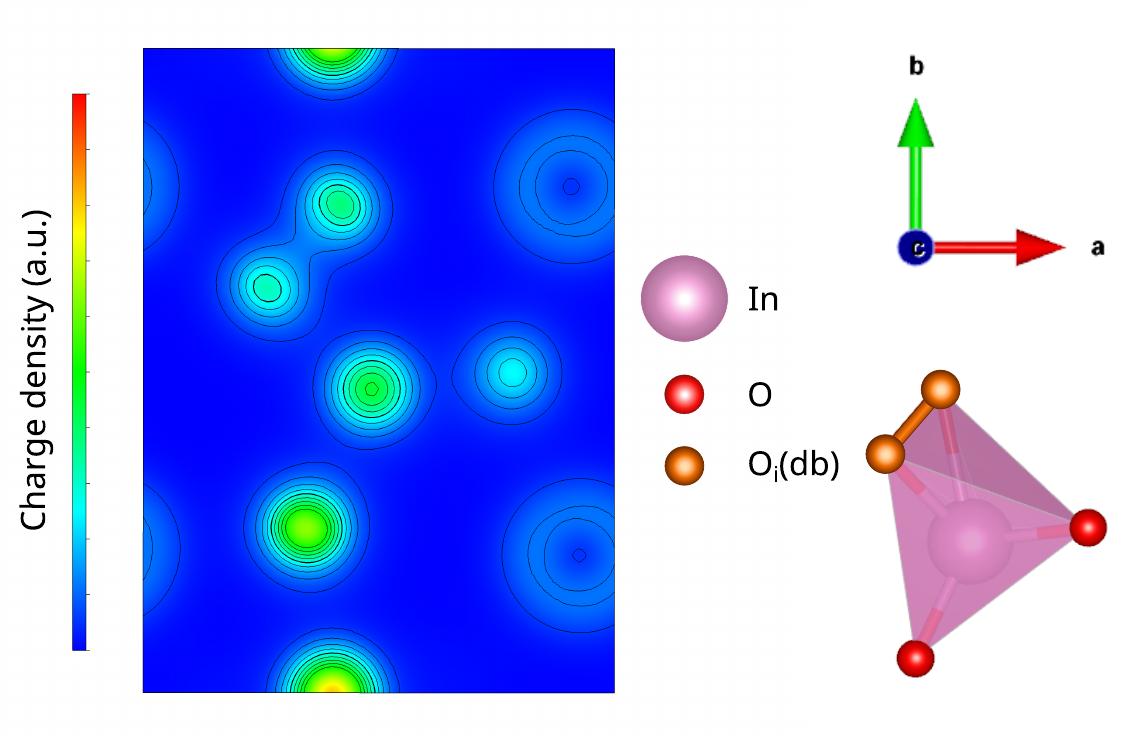}
    \caption{Structure and electronic density of dumbbell interstitial O$_i\textrm{(db)}$ configuration. The electron density iso-surface plot shows a cut parallel to the $[001]$ plane. The illustration demonstrates the strong covalent bond between the two oxygen atoms of the dumbbell.}
    \label{fig:O-db-charge}
\end{figure}

\begin{figure}
    \centering
    \includegraphics[width=\linewidth]{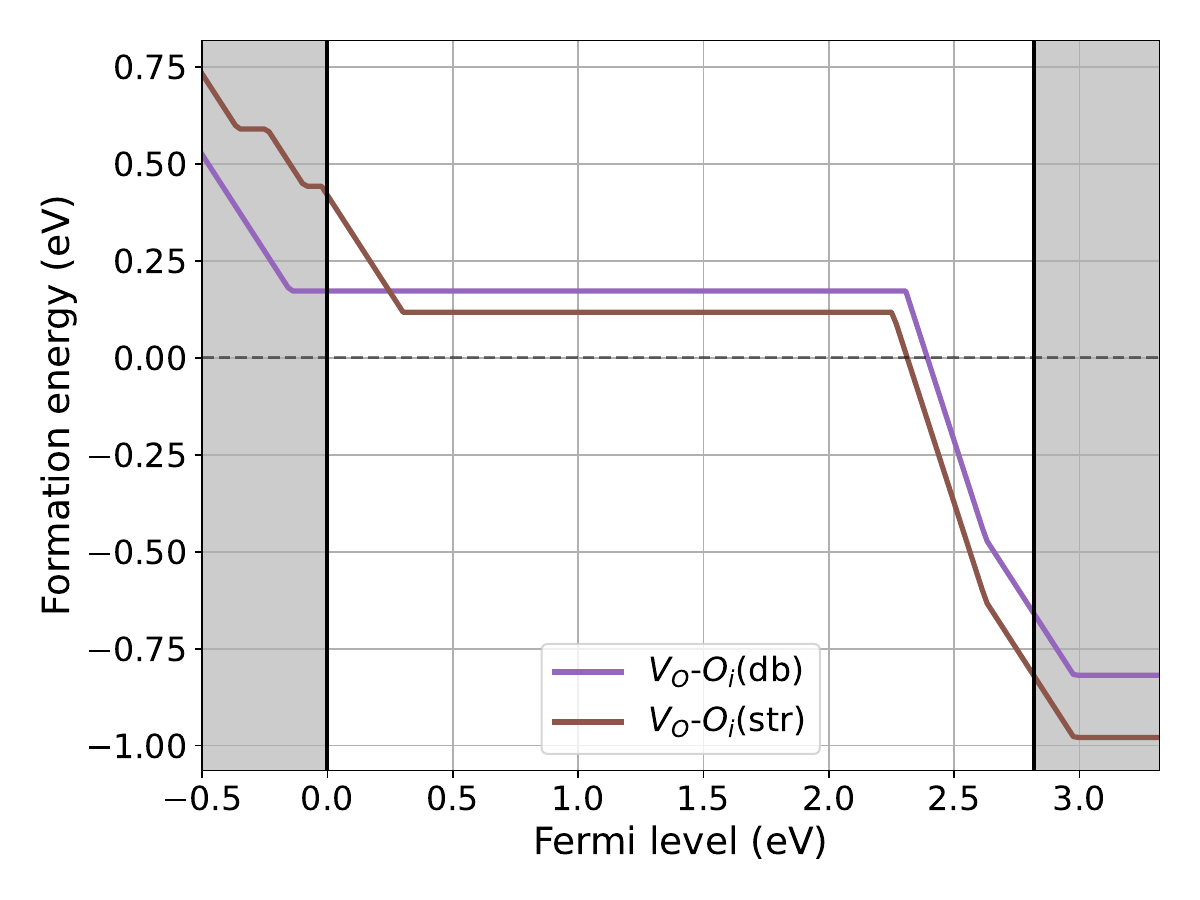}
    \caption{Binding energy of the defect complexes as a function of the Fermi level. The kinks correspond to charge transition points of the isolated defects.}
    \label{fig:Ef_binding_energies}
\end{figure}
    
O$_i$(db) presents a $+1/0$ CTL at \qty{0.15}{\eV} below the VBM. However, it remains neutral across the whole band gap. To elucidate the origin of this behavior, the electronic density of the defect was analyzed, as shown in Figure \ref{fig:O-db-charge}. The charge density reveals the formation of a covalent bond between the oxygen interstitial and a neighboring lattice oxygen, resulting in an O-O dumbbell configuration. This defect can therefore be thought of as a peroxo-ion (\ce{O2^{2-}}).
On the other hand, O$_i\textrm{(str)}$ does not form an oxygen dumbbell and presents a transition from $+1$ to $0$ at \qty{0.36}{\eV} below the VBM, another from $0$ to $-1$ at \qty{0.10}{\eV} below the VBM and another one from $-1$ to $-2$ at \qty{0.30}{\eV} above thw VBM, a much more expected behavior of a negatively-charged oxygen interstitial.
    
The \ce{Ba} vacancies are acceptors as expected, with a $0/-1$ CTL at \qty{0.10}{\eV} below the VBM and a $-1/-2$ CTL at \qty{0.41}{\eV} above it. \ce{In} vacancies also act as acceptors with a $0/-1$ CTL at \qty{0.12}{\eV} below the VBM, a $-1/-2$ CTL at \qty{0.30}{\eV} above the VBM, and a final transition from charge $-2$ to $-3$ at \qty{0.60}{\eV} above the VBM. \ce{O} vacancies are mostly donors with a $+2$ charge until \qty{0.19}{\eV} below the CBM, where it changes to a $+1$ charge state, and another transition to neutral charge \qty{0.16}{\eV} above the CBM. 
    
To analyze the behavior of the Frenkel defects, the O$_i\textrm{(db)}$ and O$_i\textrm{(str)}$ interstitials were paired with the most distant O site from each of them, which was removed to create a vacancy. The V$_{\ce{O}}$-O$_i\textrm{(db)}$ complex presents a single $+2/0$ CTL at \qty{0.51}{\eV} below the CBM, while V$_{\ce{O}}$-O$_i\textrm{(str)}$ presents a $+2/+1$ CTL at \qty{0.24}{\eV} below the VBM, a $+1/0$ CTL at \qty{0.02}{\eV} below the VBM, and a $0/-2$ CTL at \qty{0.56}{\eV} below the CBM.
    
For \ce{O}-poor conditions (reservoirs A and B) and under $p$-type conditions, there is a strong preference for the formation of both Frenkel defect complexes. Furthermore, for $n$-type conditions, \ce{O} vacancies also play a role with the O$_i\textrm{(str)}$ interstitial. Under all Fermi level values, \ce{Ba} and \ce{In} vacancies have high formation energies. They are therefore expected not to play a significant role in the defect thermodynamics of BIO.
For \ce{O}-rich conditions (reservoirs C and D) and under $p$-type conditions, O$_i\textrm{(db)}$ and its respective Frenkel pair defect are preferred, and we also notice that \ce{Ba} and \ce{In} vacancies formation energies are lowered with respect to \ce{O}-poor conditions. For $n$-type conditions, O$_i\textrm{(str)}$ as well as V$_{\ce{Ba}}$ and V$_{\ce{In}}$ become the most preferred defects.
    
We have also computed the formation energies for temperature and pressure conditions present during sintering of BIO (reservoir X in Table \ref{tab:mu_values} and Figure \ref{fig:sd}), which are shown in the bottom left panel of Figure \ref{fig:Ef_ABCDX_CTLs}.
For this reservoir and for $p$-type conditions, we observe that O$_i\textrm{(db)}$ and both Frenkel pairs have low formation energies, underlining their important contribution to charge neutrality. For $n$-type, the defects show a similar behavior to what we have described for the C and D reservoirs; however, the O$_i\textrm{(str)}$ interstitial becomes the defect with the lowest positive formation energy.
    
Figure \ref{fig:Ef_binding_energies} shows the binding energy as a function of the Fermi level for the two computed defect complexes, calculated using Equation \ref{eq:binding-energy}.
The observed changes in slope correspond to charge transitions of the constituent defects. Notably, both defect complexes become energetically favorable only above certain Fermi level thresholds: \qty{2.39}{\eV} for V$_{\ce{O}}$-O$_i$(db) and \qty{2.26}{\eV} for V$_{\ce{O}}$-O$_i$(str). Below these thresholds, the positive binding energies indicate that forming complexes is energetically unfavorable; beyond these points, complex formation leads to energy gains compared to isolated defects.

\begin{table}
    \renewcommand{\arraystretch}{1.3}
    \centering
    \caption{Distinct relaxed oxygen interstitial configurations in BIO (fractional coordinates in the relaxed PBE supercell for charge state $q=0$). For the dumbbell-type interstitials, the coordinates of both atoms occupying the interstitial site are reported.}
        \begin{tabular*}{\columnwidth}
        {@{\extracolsep{\fill}}l c c c@{}}
        \hline\hline
        Configuration & $x$ & $y$ & $z$ \\
        \hline
        \multicolumn{4}{@{\hspace{0pt}}l}{\textbf{Dumbbell-type interstitials (db)}} \\
        O$_i$(db) & 0.43552 & 0.31653 & 0.06758 \\
         & 0.50471 & 0.37584 & 0.00008 \\
        O$_i$(db$'$) & 0.34971 & 0.93937 & 0.86950 \\
        & 0.39834 & 0.01873 & 0.89332 \\[0.3em]
        \hline
        \multicolumn{4}{@{\hspace{0pt}}l}{\textbf{Structural interstitials (str)}} \\
        O$_i$(str) & 0.10129 & 0.75000 & 0.14773 \\
        \hline
        \multicolumn{4}{@{\hspace{0pt}}l}{\textbf{Other interstitials (o)}} \\
        O$_i$(o) & 0.60371 & 0.25492 & 0.86093 \\
        O$_i$(o$'$) & 0.71405 & 0.25045 & 0.93782 \\
        O$_i$(o$''$) & 0.25000 & 0.00000 & 0.11513 \\
        \hline\hline
        \end{tabular*}
    \label{tab:int_sites}
\end{table}

\subsection{Dependency on oxygen partial pressure}
\begin{figure}[t]
    \centering
        \begin{subfigure}{\linewidth}
        \centering
        \includegraphics[width=\linewidth]{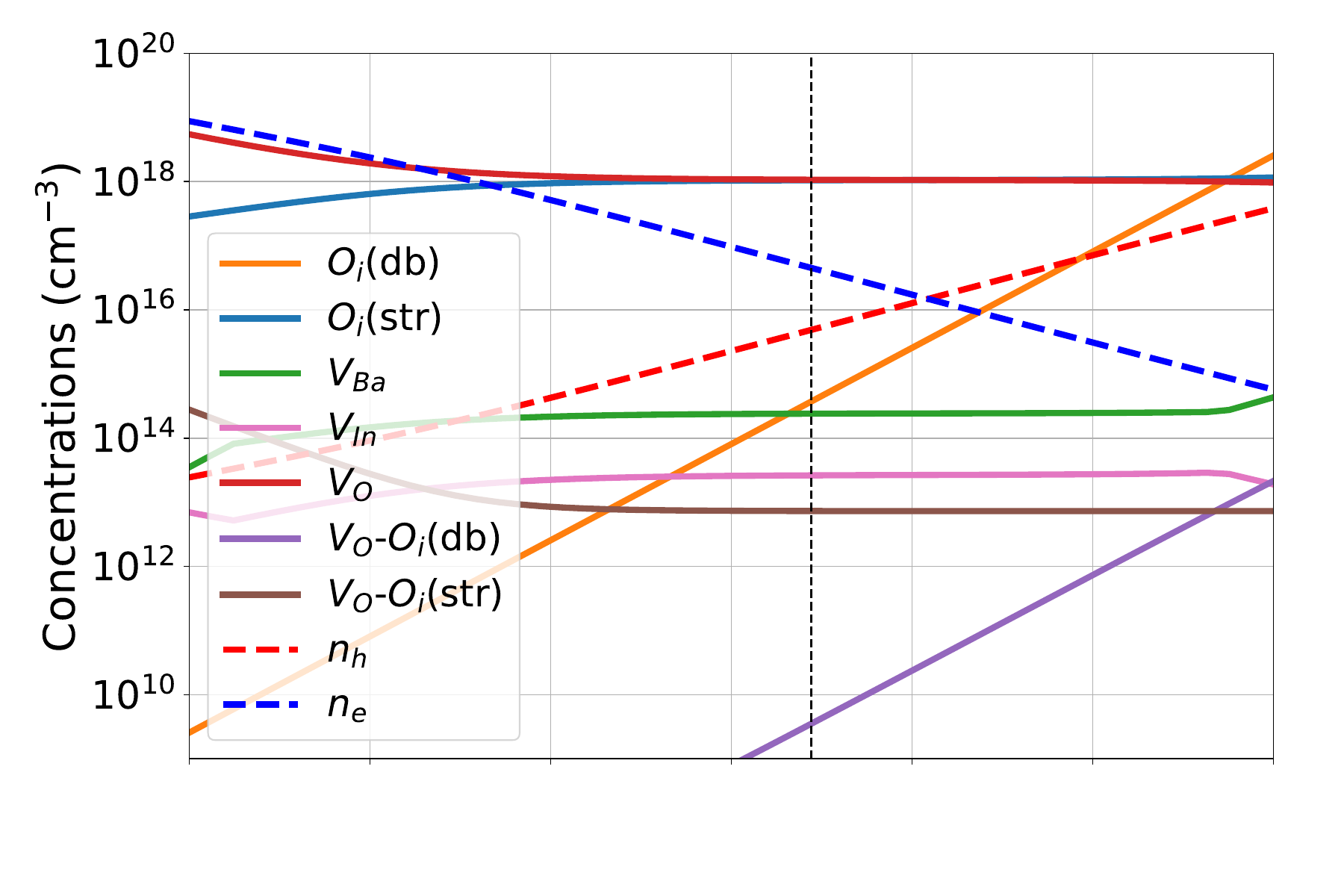}
        \end{subfigure}
        \begin{subfigure}{\linewidth}
        \centering
        \includegraphics[width=\linewidth]{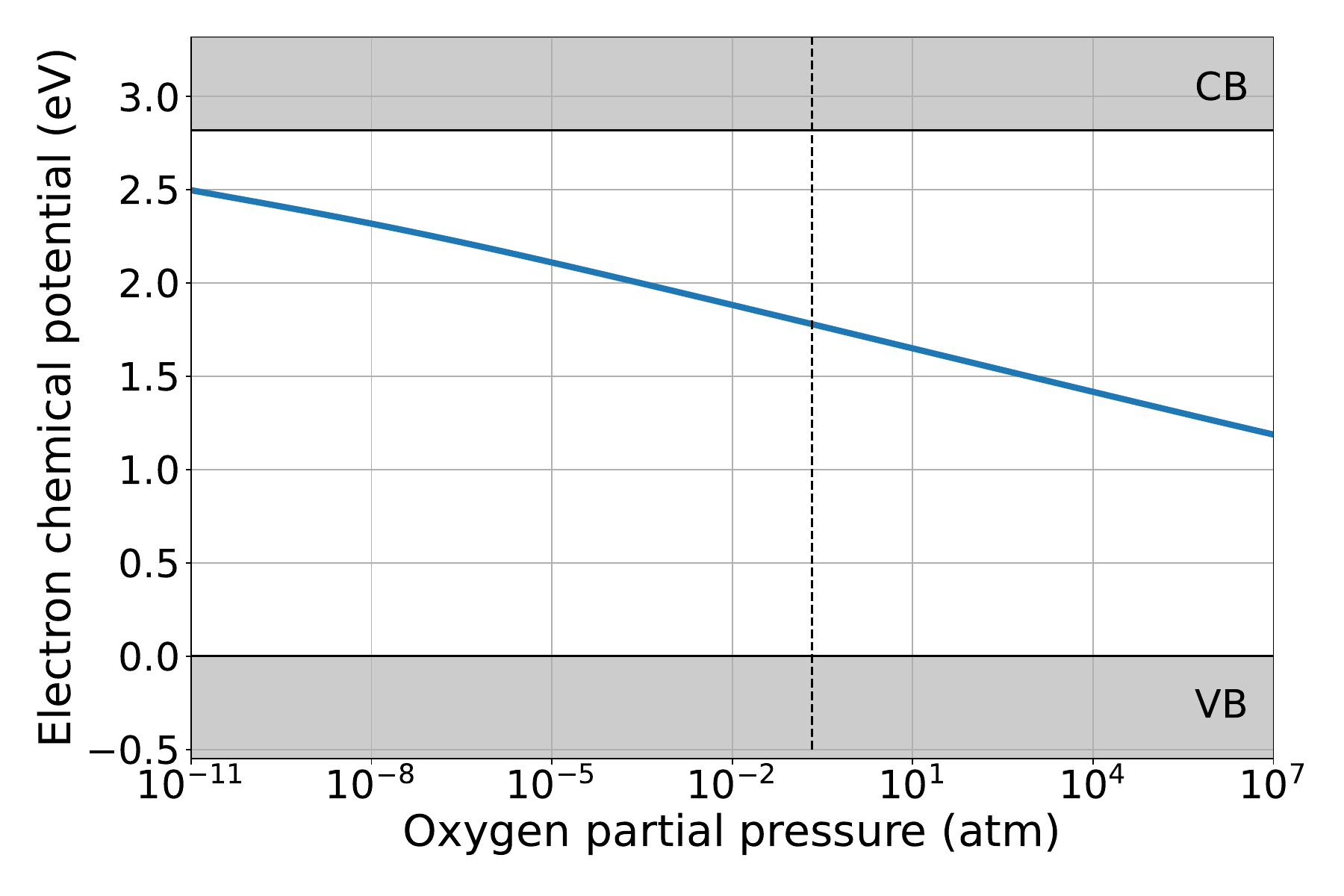}
        \end{subfigure}
    \caption{Concentrations of defects, electrons ($n_e$), and holes ($n_h$) (top), and corresponding electron chemical potential as a function
    of oxygen partial pressure (bottom) at $T = 1573\,\mathrm{K}$. The vertical dashed lines indicate atmospheric pressure of \qty{0.21}{\atm}.}
    \label{fig:brouwer_intrinsic_e_chempot}
\end{figure}

As already mentioned in Section \ref{sec:O_chem_pot}, the oxygen chemical potential can be controlled experimentally by varying the oxygen partial pressure. Therefore, it is useful to calculate the defect concentrations and the electron chemical potential as a function of the partial pressure, allowing for the correlation of computational and experimental data. To this day, there is still no reported experimental data on defect concentrations in BIO. 
We have converted the values of $\Delta\mu_{\ce{O}}$ in oxygen partial pressures using Equation \ref{eq:mu_O}, keeping the constraints of the stability over competing phases, which restricts the range of $\Delta\mu_{\ce{O}}$. For every oxygen partial pressure, we have solved the charge neutrality condition self-consistently, following Section \ref{sec:def_thermo} to find the electron chemical potential. At this point, knowing $\mu_e$, carrier and defect concentrations can be extracted. We have reported the values of the concentrations of holes ($n_h$), electrons ($n_e$), and intrinsic defect concentrations as a function of oxygen partial pressure at sintering temperature (1573 K) in the top panel of Figure \ref{fig:brouwer_intrinsic_e_chempot}.

At low $p_{\ce{O2}}$, the majority charge carriers are electrons and V$_{\ce{O}}$. With increasing $p_{\ce{O2}}$, the concentration of V$_{\ce{O}}$ decreases. In contrast, O$_i$(str) increases until both reach a concentration of \qty{e18}{\per\cubic\centi\metre}, entering the stoichiometric regime around an oxygen partial pressure of \qty{e-1}{\atm}. The stoichiometric point, at which the concentration of holes and electrons is the same, is reached at an oxygen partial pressure of \qty{e4}{\atm}. At such high $p_{\ce{O2}}$ values, the system remains in the stoichiometric regime; however, it can be observed that O$_i$(db) is present in substantial concentrations. As discussed previously, O$_i$(db) is a charge-neutral defect and therefore does not participate in charge compensation. Nevertheless, its appreciable concentration suggests that it may influence oxygen transport properties, for example, by contributing to oxygen migration pathways or modifying effective migration barriers. A quantitative assessment of this effect requires an explicit treatment of defect diffusion, which is beyond the scope of the present work.
    
The behavior of $\mu_e$ is shown in the bottom panel of Figure \ref{fig:brouwer_intrinsic_e_chempot}. For low $p_{\ce{O2}}$ values, the system shows $n$-type behavior. With increasing $p_{\ce{O2}}$, the electron chemical potential decreases until it reaches the middle of the gap.

\section{Summary and conclusions}
\label{sec:conc}
In this work, we have investigated the defect chemistry of \ce{Ba2In2O5} using density functional theory (DFT), with particular focus on vacancies, oxygen interstitials, and Frenkel pairs. Our results suggest that the intrinsic defects landscape is governed by oxygen vacancies and interstitials, whereas Ba and In vacancies present high formation energies and are therefore unlikely to contribute significantly to the defect thermodynamics. 
Among the oxygen interstitials, we observe the formation of dumbbell structures that remain neutral across the band gap. This behavior has been previously observed for oxygen interstitials in \ce{ZnO} and \ce{In2O3}. Other configurations, such as O$_i$(str), are charge carriers and become increasingly favorable under $n$-type conditions. At high oxygen partial pressures, O$_i$(str) and V$_{\ce{O}}$ dominate the defect population, suggesting that this interstitial plays a central role in the compensation mechanisms in BIO, in contrast to many other oxides where cationic vacancies usually provide charge compensation.
Additionally, diffusion studies of oxygen interstitials and vacancies would provide deeper insights into migration barriers and diffusivities, allowing for a better comparison with experimental studies. Such investigations will help to give a more comprehensive picture of the defect chemistry of BIO and its potential for defect and doping engineering in real-world applications.

\section*{Acknowledgements}
The presented work has been performed in the framework of the collaborative research centre FLAIR (Fermi level engineering applied to oxide electroceramics), which is funded by the German Research Foundation (DFG), project-ID 463184206 – SFB 1548. The authors gratefully acknowledge the computing time provided to them at the NHR Center NHR4CES at TU Darmstadt (project number p0023194). This is funded by the Federal Ministry of Research, Technology, and Space, and the state governments participating on the basis of the resolutions of the GWK for national high-performance computing at universities (\url{www.nhr-verein.de/unsere-partner}).
We would like to thank Dr.~Jochen Rohrer, Dr.~Lorenzo Villa, and Dr.~Sabrina Sicolo for helpful discussions and advice.
There are no conflicts of interest to declare.

\section*{Declaration of generative AI and AI-assisted technologies in the manuscript preparation process}

During the preparation of this work, the authors used ChatGPT-5.2 (OpenAI) for code review and editorial control. After using this tool/service, the authors reviewed and edited the content as needed and take full responsibility for the content of the published article.


\bibliographystyle{elsarticle-num-names}
\biboptions{sort&compress}
\bibliography{paper.bib}






\end{document}


\begin{frontmatter}



\title{Supplemental material for ``Defect thermodynamics of orthorhombic Ba$_2$In$_2$O$_5$:\\ First-principles calculations on the role of oxygen dumbbell interstitials''}


\author[first]{Rachele Sciotto}
\author[first]{Karsten Albe}
\affiliation[first]{organization={Department of Materials and Earth Sciences, Technical University of Darmstadt},
            addressline={Otto-Berndt-Str. 3}, 
            city={Darmstadt},
            postcode={64287}, 
            country={Germany}}

\begin{keyword}
Barium indate \sep Oxygen interstitials \sep Defect thermodynamics \sep Density functional theory \sep Charge compensation



\end{keyword}

\end{frontmatter}

\begin{figure}[h]
    \centering
    \includegraphics[width=0.5\linewidth]{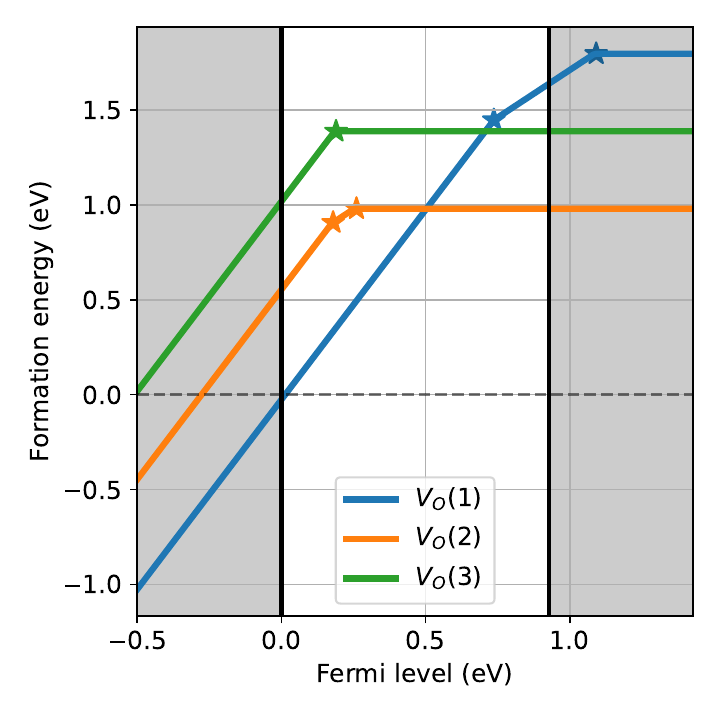}
    \caption{Defect formation energies as a function of the Fermi level position for thermodynamic conditions representative of the experimental conditions for the distinct oxygen vacancies.}
    \label{fig:Ef_O_vac}
\end{figure}
    
\begin{figure}
    \centering
    \includegraphics[width=0.5\linewidth]{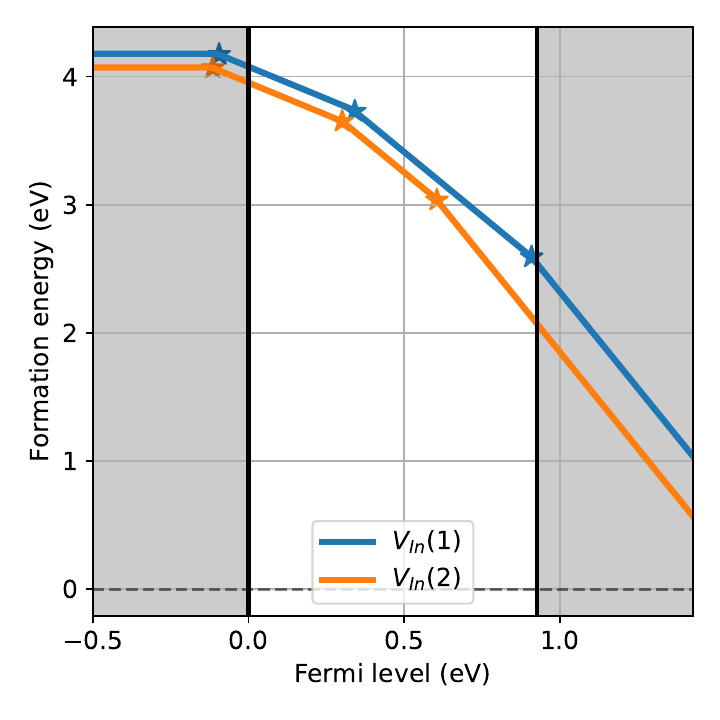}
    \caption{Defect formation energies as a function of the Fermi level position for thermodynamic conditions representative of the experimental conditions for the distinct indium vacancies.}
    \label{fig:Ef_In_vac}
\end{figure}

\begin{figure}
    \centering
    \includegraphics[width=0.5\linewidth]{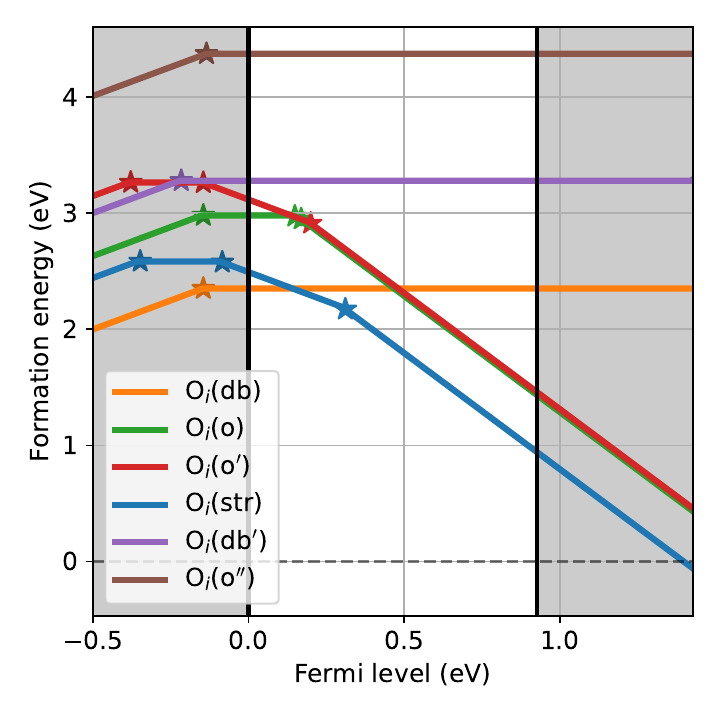}
    \caption{Defect formation energies as a function of the Fermi level position for thermodynamic conditions representative of the experimental conditions for the distinct oxygen interstitials.}
    \label{fig:Ef_O_i}
\end{figure}